\documentclass[a4paper]{jpconf}
\usepackage{graphicx}
\begin{document}
\title{Heavy flavor in relativistic heavy-ion collisions}

\author{E. L. Bratkovskaya$^{1,2}$, T. Song$^{1,2}$,
H. Berrehrah$^{1,2}$, D.  Cabrera$^{1,2}$,
J.~M. Torres-Rincon$^{3}$, L. Tolos$^{2,4}$, W. Cassing$^5$}
\address{$^1$ Institute for Theoretical Physics, JWG Universit\"{a}t, Frankfurt/M, Germany \\
$^2$ Frankfurt Institute for Advanced Studies, JWG Universit\"{a}t, Frankfurt/M, Germany\\
$^3$ Subatech, UMR 6457, IN2P3/CNRS,
  Universit$\acute{e}$ de Nantes, $\acute{E}$cole des Mines de Nantes \\ 
$^4$ Institut de Ciencies de l'Espai (IEEC/CSIC), Campus Univ. 
Autonoma de Barcelona, Spain \\ 
$^5$Institut f\"{u}r Theoretische Physik, Universit\"{a}t Gie\ss en, Germany}

\ead{Elena.Bratkovskaya@th.physik.uni-frankfurt.de}

\begin{abstract}
We study charm production in ultra-relativistic heavy-ion
collisions by using the Parton-Hadron-String Dynamics (PHSD)
transport approach.  The initial charm quarks are produced by the
PYTHIA event generator tuned to fit the transverse momentum
spectrum and rapidity distribution of charm quarks from
Fixed-Order Next-to-Leading Logarithm (FONLL) calculations. The
produced charm quarks  scatter in the quark-gluon plasma (QGP)
with the off-shell partons whose masses and widths are given by
the Dynamical Quasi-Particle Model (DQPM), which reproduces the
lattice QCD equation-of-state in thermal equilibrium. The relevant
cross sections are calculated in a consistent way by employing the
effective propagators and couplings from the DQPM.  Close to the
critical energy density of the phase transition, the charm quarks
are hadronized into $D$ mesons through coalescence and/or
fragmentation. The hadronized $D$ mesons then interact with the
various hadrons in the hadronic phase with cross sections
calculated in an effective lagrangian approach with heavy-quark
spin symmetry. The nuclear modification factor
$R_{AA}$ and the elliptic flow $v_2$ of $D^0$ mesons from PHSD are
compared with the experimental data from the STAR Collaboration
for Au+Au collisions at $\sqrt{s_{NN}}$ =200 GeV
and to the ALICE data for Pb+Pb collisions at  $\sqrt{s_{NN}}$ =2.76 TeV.
We find that in the PHSD the energy loss of $D$ mesons at high $p_T$ can be
dominantly attributed to partonic scattering while the actual
shape of $R_{AA}$ versus $p_T$ reflects the  heavy-quark
hadronization scenario, i.e. coalescence versus fragmentation.
Also the hadronic rescattering is important for the $R_{AA}$
at low $p_T$ and enhances the $D$-meson elliptic flow $v_2$.
\end{abstract}

\section{Introduction}
The Quantum Chromo Dynamics (QCD) \cite{lQCD,Borsa,Peter,Ding} predicts that matter changes its phase at high
temperature and density and bound (colorless) hadrons dissolve to interacting (colored) quarks and gluons, i.e. to the Quark-Gluon-Plasma (QGP). Such extreme conditions have existed in the early expansion of the universe and now can be realized in the laboratory by collisions of
heavy-ions at ultra-relativistic energies. The study of the phase boundary and the properties of the QGP are the main goal of several
present and future heavy-ion experiments at SPS (Super Proton Synchrotron), RHIC (Relativistic Heavy-Ion Collider), LHC (Large Hadron
Collider) and the future FAIR (Facility for Antiproton and Ion Research) and NICA (Nuclotron-based Ion Collider  fAcility)
\cite{QM2014}.  Since the QGP is created only for a short time (of a couple of fm/c) it is quite challenging to study its properties and to find the most sensible probes. The advantage of
 'hard probes' such as  mesons containing  heavy quarks (charm and beauty) is, firstly, that due to the heavy masses they are
dominantly produced in the very early stages of the reactions with large energy-momentum transfer, contrary to the light hadrons and
electromagnetic probes.  Secondly, they are not in an equilibrium with the surrounding matter due to smaller interaction cross sections
relative to the light quarks and, thus, may provide an information on their creation mechanisms. Moreover, due to the hard scale,
perturbative QCD (pQCD) should be applicable for the calculation of heavy quark production. As shown in Ref. \cite{Vogt}, the FONLL calculations
are in good agreement with the experimental observables on charm meson spectra in p+p collisions.  This
provides a solid reference frame for studying the heavy-meson production and their flow pattern in heavy-ion collisions.

The first charm measurements at RHIC energies by the PHENIX\cite{eePHENIX} and STAR \cite{eeSTAR} collaborations were
related to the single non-photonic electrons emitted from the decay of charm mesons. However, recently the STAR Collaboration measured
directly the nuclear modification factor and the elliptic flow of $D^0$ mesons in Au+Au collisions at $\sqrt{s_{\rm NN}}=$200
GeV~\cite{Adamczyk:2014uip,Tlusty:2012ix} which allows for a straightforward comparison with the theoretical model calculations. It has been
observed that the $\rm R_{AA}$ and $v_2$ of charm mesons show a similar behavior as in case of light hadrons contrary to expectations from pQCD. Similar observations have been made at LHC energies, too \cite{expLHC}.

It  still remains a challenge for the theory to reproduce the experimental data and to explain simultaneously the large energy loss of charm quarks ($\rm R_{AA}$) and the strong collectivity ($v_2$) (cf. e.g. \cite{vanHees:2005wb,BAMPS,Das:2013kea,Ozvenchuk:2014rpa}).
Commonly, the interactions of charm quarks with the partonic medium are based on pQCD with massless light quarks and a fixed or running coupling. However, lattice QCD and quasiparticle approaches lead to the notion of massive degrees-of-freedom with finite spectral width \cite{Cassing:2008nn}. In this case the pQCD cross sections are no longer meaningful. Moreover, the conclusions on the amount of suppression due to collisional energy loss by means of the elastic interactions of charm
quarks with the QGP partons versus the radiative energy loss due to the emission of soft gluons (i.e. gluon bremsstrahlung) are still far from
being robust.  Also the influence of hadronization and especially hadronic rescattering is not yet settled, too.

Our goal here is to study the charm dynamics based on a consistent microscopic transport approach for the charm production, hadronization
and rescattering with the partonic and hadronic medium. In this study we will confront our calculations within the
Parton-Hadron-String Dynamics (PHSD) approach to the experimental data on charm at RHIC and LHC energies and discuss the perspectives/problems of
using the charm quarks for the tomography of the QGP.  To achieve this goal we embed the heavy-quark physics in the existing  PHSD transport
approach \cite{PHSD} which incorporates explicit partonic degrees-of-freedom in terms of strongly interacting quasiparticles
(quarks and gluons) in line with an equation-of-state from lattice QCD (lQCD)
as well as dynamical hadronization and hadronic elastic and inelastic collisions in
the final reaction phase. Since PHSD has been successfully applied to
describe the final distribution of mesons (with light quark content)
from lower SPS up to LHC energies \cite{PHSD,PHSDrhic,Volo,Linnyk}, it
provides a solid framework for the description of the creation,
expansion and hadronization of the QGP as well as the hadronic
expansion with which the heavy quarks interact either as quarks or as
bound states such as $D$-mesons. We note in passing that in the hadronic phase PHSD merges with the familiar
Hadron-String-Dynamics (HSD) approach \cite{HSD,HSD2}.

\section{Description of charm production and dynamics in PHSD}
\subsection{Charm production in p+p collisions}
Before studying the charm production in relativistic heavy-ion
collisions, we discuss the  charm production in p+p collisions at the
top RHIC energy of $\sqrt{s_{\rm NN}}$ = 200 GeV.  The charm production in
p+p collisions also plays the role as a reference for the nuclear
modification factor $\rm R_{AA}$ in heavy-ion collisions.

\begin{figure}[t]
\phantom{a} \vspace*{-10mm}
\centerline{
\includegraphics[width=8.8 cm]{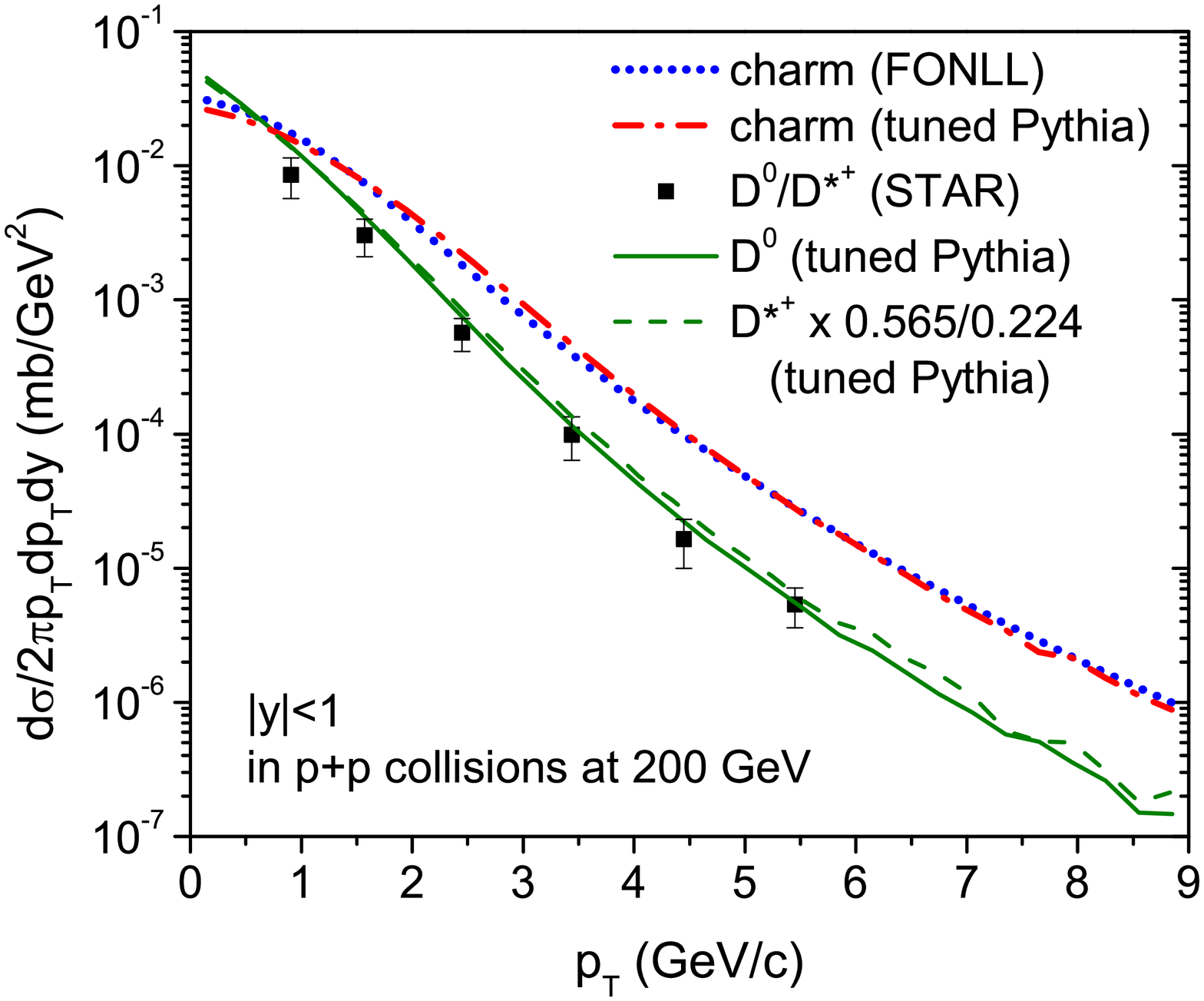} \includegraphics[width=8.7 cm]{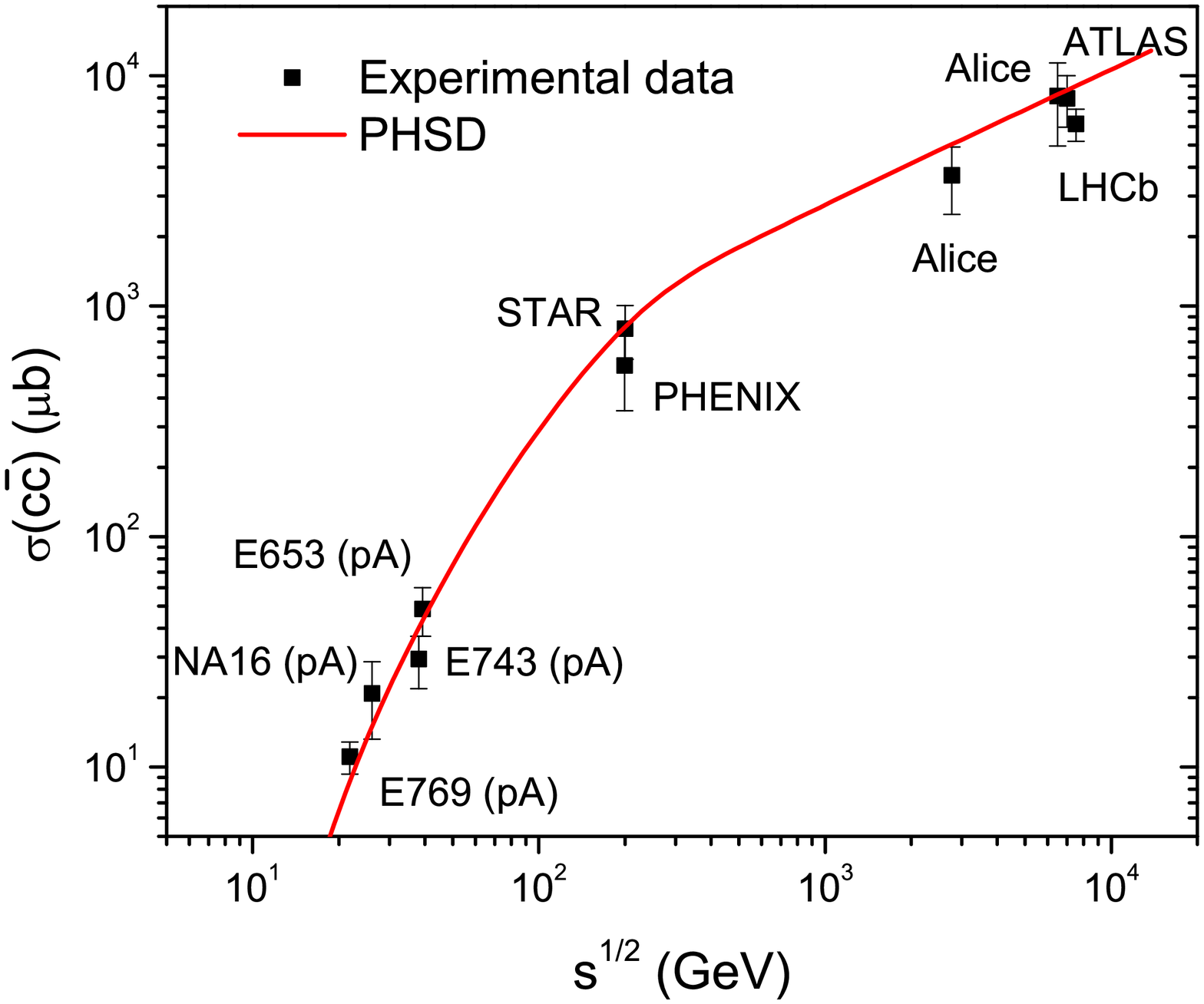}}
\caption{(l.h.s.) Transverse momentum
distributions (upper lines) of charm quarks in p+p collisions at $\sqrt{s_{\rm NN}}=$200 GeV from FONLL (dotted lines) and the tuned Pythia event generator (dot-dashed lines); in the lower lines the transverse momentum spectrum of $D^0$ mesons -- which are
fragmented from charm quarks with the contribution from $D^*$ decay included (solid line) --  and that of $D^{*+}$ after scaling are compared with the experimental data from the STAR Collaboration~\cite{Adamczyk:2012af}. (r.h.s.)
The total cross section for charm production in
p+p collisions (as parameterized in the PHSD) in comparison with the
experimental data at various collision energies~\cite{Adamczyk:2012af,delValle:2011ex}. }
\label{pp1}
\end{figure}

We use the Pythia event generator to produce charm and anticharm quarks
in p+p collisions with the parameters PARP(91)=1.0 GeV/c and
PARP(67)=1.0 as in Ref.~\cite{TSong}. The
former parameter is the Gaussian width of the primordial transverse
momentum of a parton which initiates a shower in
hadrons and the latter the parton shower level parameter. We note that
by an additional suppression of the transverse momenta of charm and
anticharm quarks by 10 \% and a suppression of rapidities by 16 \%  the
transverse momentum spectrum as well as rapidity distribution of charm
and anticharm quarks from the Pythia event generator are very similar
to those from the FONLL calculations in p+p collisions at $\sqrt{s_{\rm
NN}}=$200 GeV as shown in Fig.~\ref{pp1} (l.h.s.). Here the
red dot-dashed lines are from the Pythia event generator (after tuning) and the blue dotted lines from the
FONLL calculations, respectively.

The produced charm and anticharm quarks  hadronize by emitting soft
gluons.  The probabilities for a charm quark to  hadronize into
$D^0,~D^+,~D^{*0},~D^{*+},~D_s^+$, and $\Lambda_c$ are, respectively, are specified in Ref. \cite{TSong}
together with  the fragmentation function.  The solid line in Fig.~\ref{pp1} (l.h.s.) shows the transverse momentum
spectrum of $D^0$ mesons after charm quark fragmentation including the
contribution from the decay of $D^{*0}$ and $D^{*+}$. We can see that
our results reproduce the experimental data from the STAR
Collaboration  \cite{Adamczyk:2012af} reasonably well.

Since charm-quark production requires a high energy-momentum transfer,
the number of produced charm quark pairs in relativistic heavy-ion
collisions is proportional to the number of binary nucleon-nucleon
collisions $N_{bin}$. Since the probability to produce a charm quark
pair depends on invariant energy and is less than that for primary hard
collision in the PHSD, the binary nucleon-nucleon collisions producing
charm quark pairs in PHSD are chosen by Monte Carlo from the ratio
of the cross section for charm production in nucleon-nucleon collisions,
$\sigma_{c\bar{c}}^{pp}(\sqrt{s})$, to the inelastic nucleon-nucleon cross section.
The actual charm cross section $\sigma_{c\bar{c}}^{pp}(\sqrt{s})$ (red solid line) is shown in Fig. 1 (r.h.s.)
in comparison to the experimental data from \cite{Adamczyk:2012af,delValle:2011ex}.

\subsection{Heavy quark scatteruing in the QGP}
Quarks, antiquarks, and gluons are dressed in the QGP and have temperature-dependent effective masses and widths. In the DQPM, the mass and width of the light partons are given by thermal quantum-field theory assuming leading order diagrams but the strong coupling $g^2(T)$ is fitted to lattice data on energy and entropy densities, etc.~\cite{Cassing:2008nn}. Note that a nonzero width of a parton reflects the off-shell nature as well as the strong interaction of the quasi-particle and finite life-time.

The charm and anticharm quarks produced in early hard collisions interact with the dressed off-shell partons in the QGP. The cross sections for the charm quark scattering with massive off-shell partons are calculated considering the mass spectra of final state particles~\cite{Berrehrah:2013mua,Berrehrah:2014kba}. In the current study the charm quark mass is taken to be 1.5 GeV and its mass spectrum is neglected for simplicity.

In the present study we consider only elastic scattering of charm quarks by light quarks and gluons. We do not consider yet the radiative processes which generate radiative energy loss because we expect that, due to the large gluon mass in the DQPM, the radiative processes are sub-dominant as compared to the collisional ones, especially for low charm-quark momenta $(p_T)$. We expect the radiative energy loss to contribute at very high $p_T$ as accessible experimentally at the LHC~\cite{Younus:2013rja}.

We emphasize that the transport coefficient $D_s$ for charm quark diffusion -- extracted from our microscopic calculations -- and its agreement with the lQCD results (within errors) and the corresponding $D$ meson $D_s$ in hadronic medium validate our description for the coupling of charm with the QGP matter \cite{Hamza14}.

\subsection{Hadronization of charm quarks}\label{tc}

Since the hot and dense  matter created by a relativistic heavy-ion
collision expands with time, the energy density of the matter
decreases and the deconfined degrees-of-freedom hadronize to color
neutral hadronic states. Once the local energy density in PHSD
becomes lower than 0.5 ${\rm GeV/fm^3}$, the partons are
hadronized. First we look for all combinations of a charm quark
and light antiquarks or of an anticharm quark and a light quark and
calculate the probability for each combination to form a $D$ or
$D^*$ (or $D_s, D^*_s$) meson. The probability for a quark
and an anti-quark to form a meson is given by
\begin{eqnarray}
f(\vec\rho,{\bf k}_\rho)=\frac{8g_M}{6^2}
\exp\left[-\frac{\vec\rho^2}{\delta^2}-{\bf k}_\rho^2\delta^2\right],
\label{meson}
\end{eqnarray}
where $g_M$ is the degeneracy of the meson $M$, and
\begin{eqnarray}
\vec\rho=\frac{1}{\sqrt{2}}({\bf r}_1-{\bf r}_2),\quad{\bf k}_\rho=\sqrt{2}~\frac{m_2{\bf k}_1-m_1{\bf k}_2}{m_1+m_2},
\label{coalescence}
\end{eqnarray}
with $m_i$, ${\bf r}_i$ and ${\bf k}_i$ being the mass, position and momentum of the quark or antiquark $i$, respectively. The width parameter $\delta$ is related to the root-mean-square radius of the meson produced through
$\langle r^2 \rangle=3(m_1^2+m_2^2)/(2(m_1+m_2)^2) \ \delta^2
$ and thus determined by experiment (if available). However, we will use this radius as a free parameter and evaluate
the coalescence for a $D$-meson radius of 0.5 and 0.9 fm (see below). For further details we refer the reader to  \cite{TSong}.

Collecting all possible combinations of a charm or an anticharm quark
with light quarks or antiquarks and calculating the coalescence
probability for each combination from Eq.~(\ref{meson}), we
obtain the probability for the charm or the anticharm quark to
hadronize by coalescence in the actual space-time volume
$\Delta t \Delta x \Delta y \Delta z$.  Whether the
charm or the anticharm quark is actually hadronized by coalescence is decided by
Monte Carlo.  Once the charm or the anticharm quark is decided to be
hadronized by coalescence, then we find its partner again by Monte
Carlo on the  basis on the probability of each combination in the
selected local ensemble. In case the charm or anticharm quark is
decided not to hadronize by coalescence, it is hadronized by the
fragmentation method as in p+p collisions (cf. Section 2.1). Since the
hadronization by coalescence is absent in p+p collisions, it can be
interpreted as a nuclear matter effect on the hadronization of charm and anticharm quarks.

\subsection{$D$ meson scattering in the hadron gas}\label{HG2}

The $D$ and $D^*$ mesons produced through coalescence or
fragmentation interact with the surrounding hadrons in PHSD. The
presence of several resonances close to threshold energies with
dominant decay modes involving open-charm mesons and light hadrons
suggests that the scattering cross
sections of a $D/D^*$ off a meson or baryon, highly abundant in the
post-hadronization medium, could manifest a non-trivial energy,
isospin and flavor dependence. An example of these states is the
broad scalar resonance $D_0(2400)$, which decays into the
pseudoscalar ground state $D$ by emitting a pion in the $s$-wave
(similarly to the heavy-quark spin partner $D_1(2420)$, decaying into $D^*\pi$).
Moreover, the similarity  between the $\Lambda(1405)$ and the
$\Lambda_c(2595)$ has driven the attention to the fact that the
latter could be playing the role of a sub-threshold resonance in the
$DN$ system, connected to the latter by coupled-channel dynamics.

All these features have been addressed within several recent approaches based on hadronic effective models which incorporate chiral symmetry breaking in the light sector. The additional freedom stemming from the coupling to heavy-flavored mesons is constrained by imposing heavy-quark spin symmetry (HQSS).  Whereas chiral symmetry fully determines the scattering amplitudes of Goldstone bosons with other hadrons at leading order in a model independent way, by means of HQSS the dynamics of the pseudoscalar and the vector mesons containing heavy quarks can be connected, since all kinds of spin interactions are suppressed in the limit of infinite quark masses \cite{Tolos:2013kva}. For further details we refer the reader to Ref. \cite{TSong} and the original literature cited therein.

\section{Comparison to experiment}
We now turn to actual nucleus-nucleus collisions at RHIC and LHC energies.
Here the nuclear modification of $D$ mesons is expressed in term of the ratio $\rm R_{AA}$ which is defined as
\begin{eqnarray}
{\rm R_{AA}}(p_T)\equiv\frac{dN_D^{\rm Au+Au}/dp_T}{N_{\rm binary}^{\rm Au+Au}\times dN_D^{\rm p+p}/dp_T},
\label{raa}
\end{eqnarray}
where $N_D^{\rm Au+Au}$ and $N_D^{\rm p+p}$ are, respectively, the number of
$D$ mesons produced in Au+Au collisions and that in p+p collisions, and
$N_{\rm binary}^{\rm Au+Au}$ is the number of binary nucleon-nucleon
collisions in Au+Au (Pb+Pb) collisions for the centrality class considered. If
the matter produced in relativistic heavy-ion collisions does not
modify the $D$ meson production and propagation, the numerator of
Eq.~(\ref{raa}) should be equal to the denominator. Therefore, an $\rm
R_{AA}$ smaller or larger than one implies that the nuclear matter
suppresses or enhances $D$ mesons, respectively.
 The elliptic anisotropy  in the azimuthal angle $\psi$ is  characterized by
\begin{equation}
\label{eqv2}
 v_2 = \left<\cos(2\psi-2\Psi_{RP})\right>= \left<\frac{p^2_x - p^2_y}{p^2_x + p^2_y}\right>~,
\end{equation}
where $p_x$ and $p_y$ are the $x$ and $y$ components of the particle momenta and $\Psi_{RP}$ is the azimuth of the reaction plane while the brackets denote averaging over particles and events. The $v_2$  coefficient can be considered as a function of centrality, rapidity $y$ and/or transverse momentum $p_T$.  We note that the reaction plane in PHSD is given by the $(x - z)$ plane with the $z$-axis in the beam direction.

\begin{figure}[tbh]
\vspace*{-5mm}
\begin{minipage}[l]{7.5cm}
\includegraphics[width=8.5 cm]{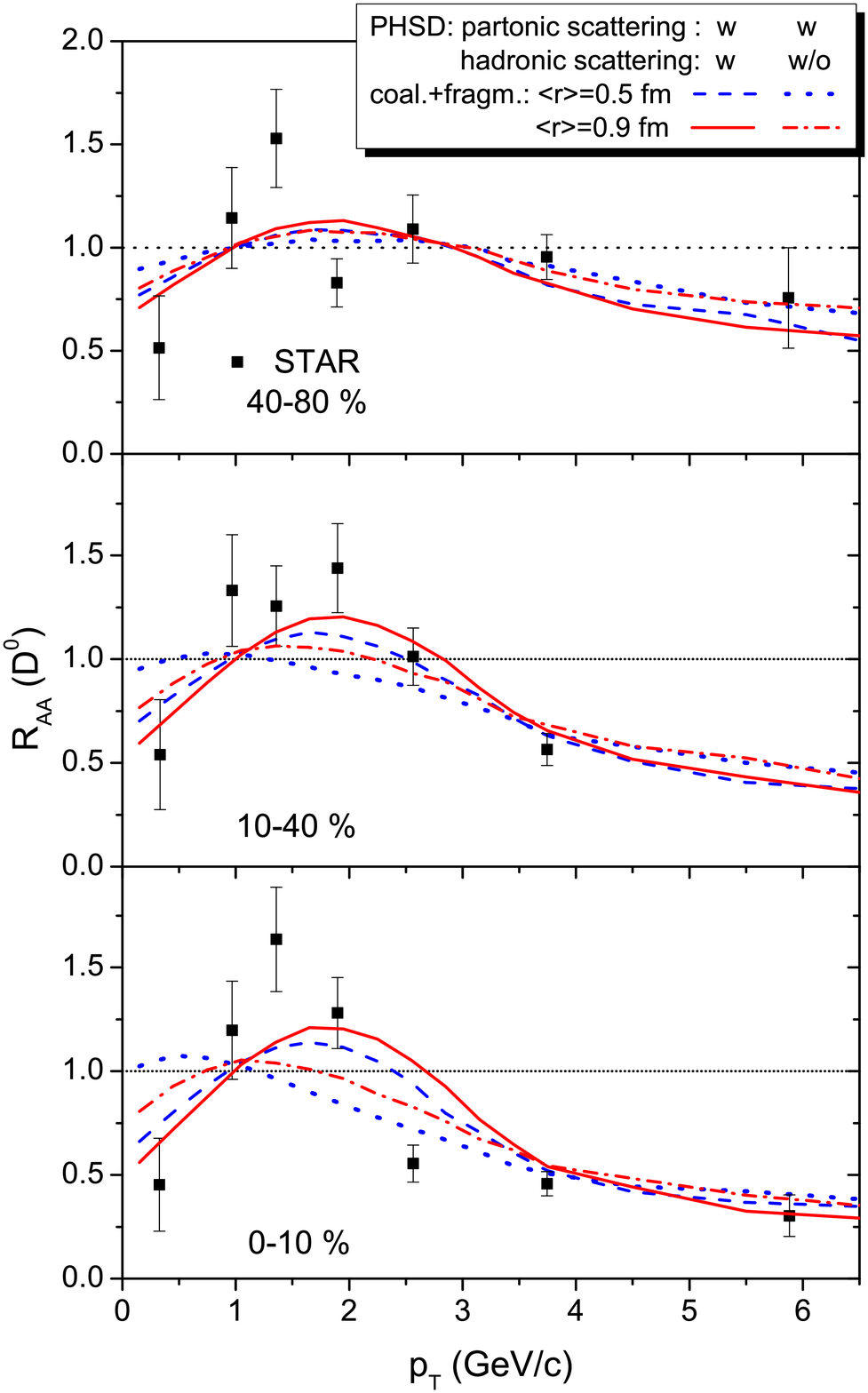}
\end{minipage}\hfill
\begin{minipage}[r]{7.5cm}
\includegraphics[width=8.5 cm]{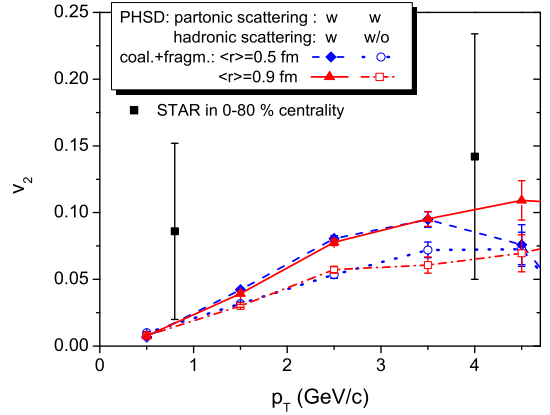}
\caption{(l.h.s.) The $\rm R_{AA}$  of
$D^0$ mesons including partonic scattering with (dashed and solid lines) and without hadronic scattering (dotted and dot-dashed lines) in Au+Au collisions at $\sqrt{s_{\rm NN}}=$200 GeV for a $D$ meson
radius of 0.5 fm and of 0.9 fm. The experimental data are from the
STAR Collaboration~\cite{Adamczyk:2014uip,Tlusty:2012ix}. (r.h.s.)
The elliptic flow $v_2$ of $D^0$ mesons including partonic scattering with
(dashed and solid lines) and without hadronic scattering (dotted and
dot-dashed lines) in Au+Au collisions at
$\sqrt{s_{\rm NN}}=$200 GeV for a $D$ meson radius of 0.5
fm and of 0.9 fm.}
\label{hadrons}
\end{minipage}
\end{figure}

Fig.~\ref{hadrons} shows the $\rm R_{AA}$ (l.h.s.) and  the elliptic flow $v_2$ of
$D^0$ mesons  (r.h.s.) with and without hadronic scattering. We can see that the
hadronic scattering plays an important role both in $\rm R_{AA}$ and
the elliptc flow $v_2$. It shifts the peak of $\rm R_{AA}$ to higher
transverse momentum especially in central collisions and enhances the
elliptic flow of final $D$ mesons.

\begin{figure}[tbh]
\label{RAA_LHC}
\centerline{
\includegraphics[width=5.8 cm,angle=270]{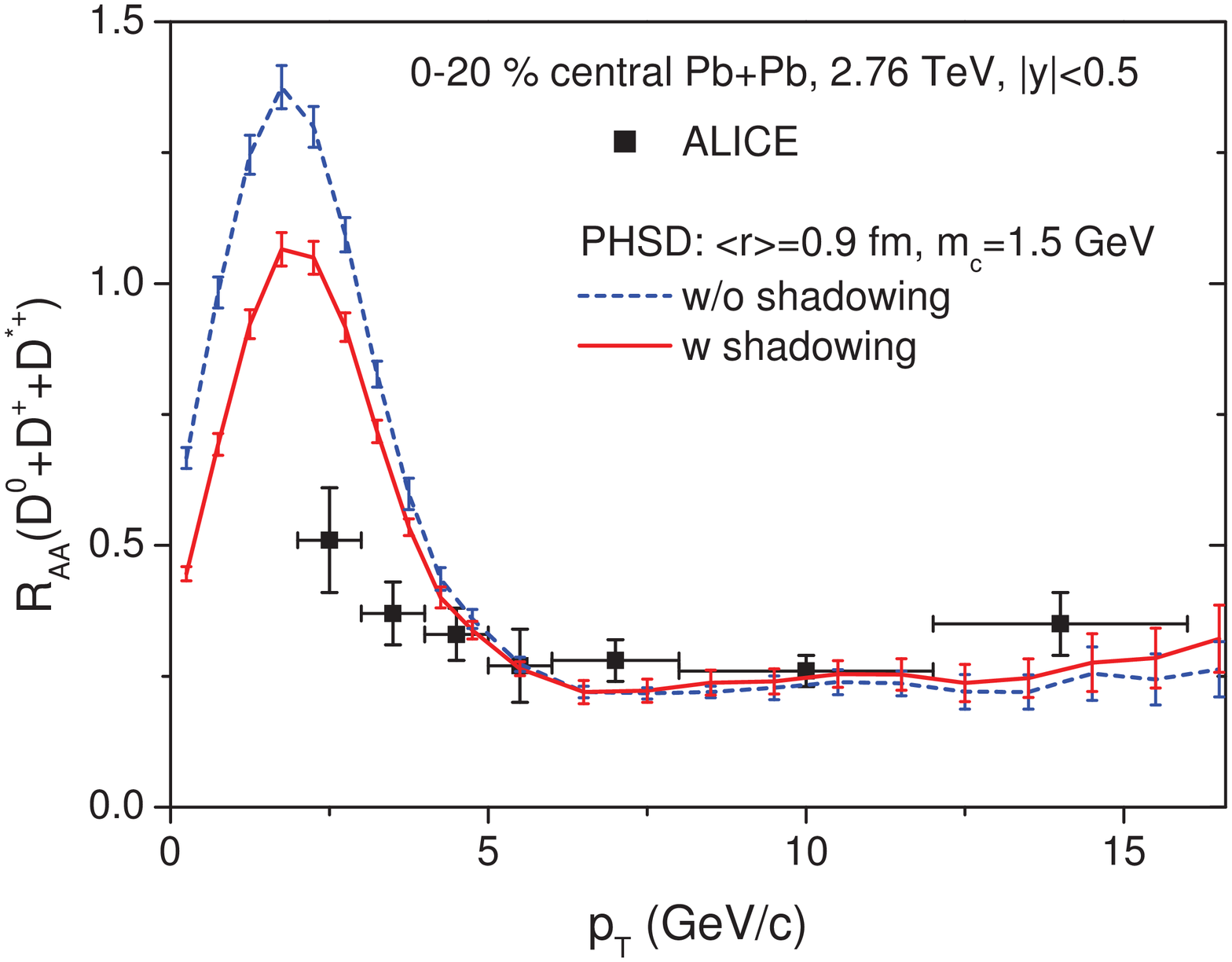}
\hspace*{3mm}\includegraphics[width=5.8 cm,angle=270]{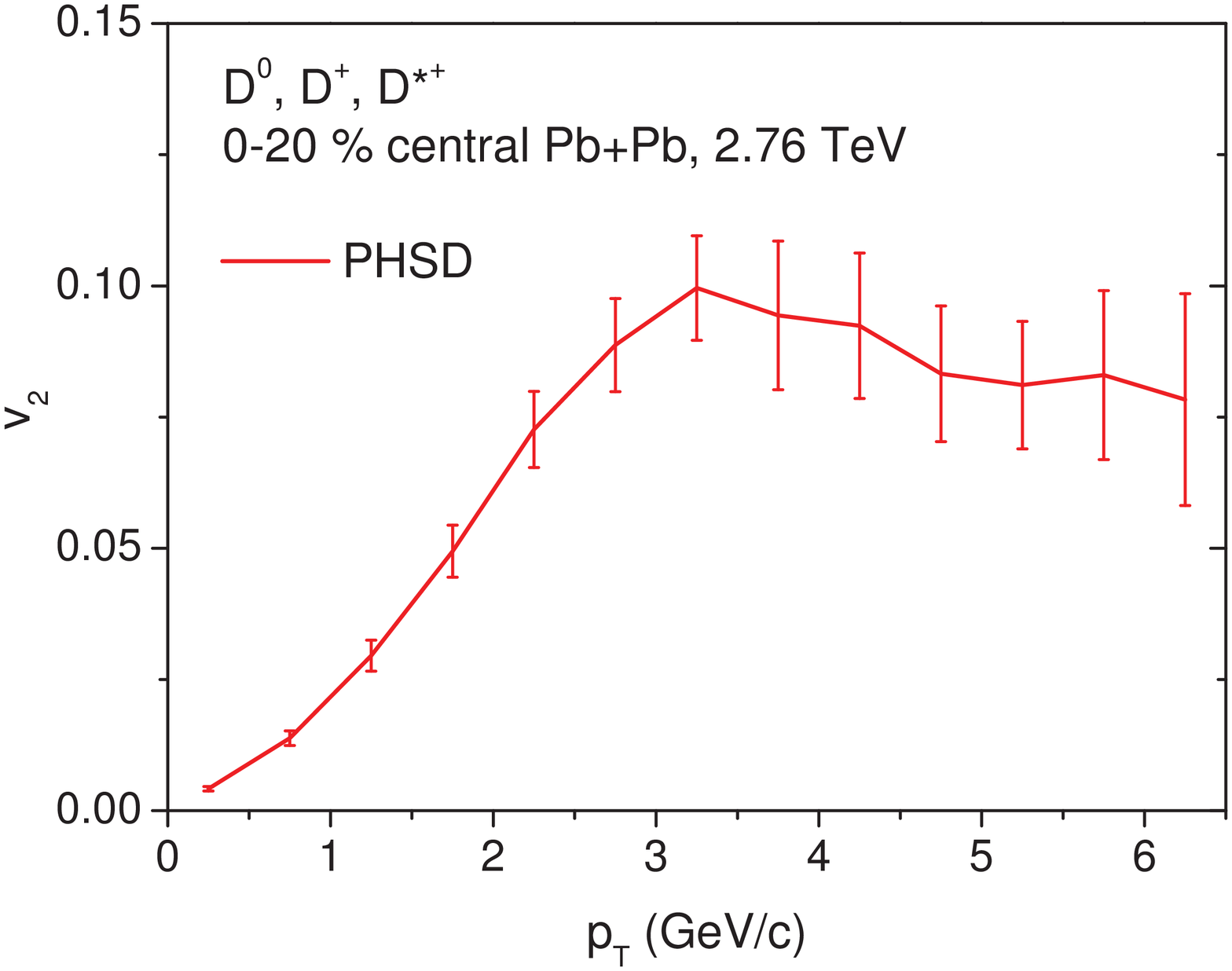} }
\caption{
(l.h.s.) The PHSD results for the $\rm R_{AA}$
of $D^0+D^+ +D^{*+}$ mesons at midrapidity for 0-20\% central Pb+Pb
collisions at $\sqrt{s_{\rm NN}}=$2.6 TeV calculated
without (dashed blue line) and with (solid red line) shadowing
in comparison to the experimental data from the ALICE
Collaboration~\cite{ALICE:2012ab}.
(r.h.s.) The PHSD results for the $v_2$ versus $p_T$ of
$D^0+D^+ +D^{*+}$ mesons for 20\% central Pb+Pb collisions at
$\sqrt{s_{\rm NN}}=$2.6 TeV including the shadowing effect.}
\end{figure}

In Fig. \ref{RAA_LHC}  we present our {\em preliminary} results for the LHC energies.
The left part of the figure  shows the PHSD results for
the $\rm R_{AA}$ of $D^0+D^+ +D^{*+}$ mesons at midrapidity
for 20\% central Pb+Pb collisions at $\sqrt{s_{\rm NN}}=$2.76 TeV
calculated without (red line) and with (blue line) shadowing
\cite{Shadowing} in comparison to the experimental data from the ALICE
Collaboration~\cite{ALICE:2012ab}.
As seen the shadowing effect reduces the peak  at low $p_T$,
 which is still higher than the experimental data.
The suppression of the high $p_T$ tail is comparable with the corresponding
results for RHIC energies and approximately in line with the data.
We stress again that we did not account for radiative energy loss
in the present calculations which might play a role
at higher $p_T$ \cite{Younus:2013rja}.

The right part of figure \ref{RAA_LHC} shows the corresponding
$v_2$ for 20\% central Pb+Pb collisions for the scenario
including shadowing.
We note that the $v_2$-flow for central (0-20\%) collisions is lower
then that for semi-central (30-50\%) measured by the ALICE
Collaboration \cite{ALICE:2013v2}, however, still quite substantial (up to 10\%)
and comparable with the $v_2$ of light hadrons for the same centrality.

\section{Summary}\label{summary}

We have studied charm production in relativistic heavy-ion
collisions by using the Parton-Hadron-String Dynamics (PHSD) approach
\cite{PHSDrhic} where the initial charm quark pairs are produced in binary
nucleon-nucleon collisions from the PYTHIA event generator  taking into
account the smearing of the collision energy due to the Fermi motion of
nucleons in the initial nuclei.  The produced charm and anticharm
quarks interact with the dressed quarks and gluons in the QGP which are
described by the Dynamical Quasi-Particle Model \cite{Cassing:2008nn} in PHSD. The interactions of
the charm quarks with the QGP partons have been evaluated with the DQPM
propagators and couplings consistently~\cite{Berrehrah:2013mua}.
We mention that when extracting the spatial diffusion coefficient $D_s$ from our
cross sections as a function of the temperature we observe
a minimum of $D_s$ close to $T_c$ which is in line with lattice data above $T_c$
and hadronic many-body calculations below $T_c$ (cf. Ref. \cite{Hamza14}).

We have found that the
interaction of the charm quarks with the dynamical partons of the QGP softens the $p_T$ spectrum of charm and
anticharm quarks but does not lead to a full thermalization for transverse
momenta $p_T >$ 2 GeV/c.  The charm
and anticharm quarks, furthermore,  are hadronized to $D$ mesons either through the
coalescence with a light quark or antiquark or through the
fragmentation by emitting soft 'perturbative' gluons.
Since the hadronization through
coalescence is absent in p+p collisions, it can be interpreted as a
nuclear matter effect on the $D$ meson production in relativistic
heavy-ion collisions.  In the coalescence mechanism the charm or
anticharm quark gains momentum by fusing with a light quark or
antiquark while it looses momentum in the fragmentation process (as in
p+p reactions).  This partly contributes to the large $\rm R_{AA}$ of
$D$ mesons between 1 and 2 GeV/c of transverse momentum.  Finally, the
formed $D$ mesons interact with hadrons by using the cross sections
calculated in an effective lagrangian approach with heavy quark
spin-symmetry \cite{Tolos:2013kva}, which is state-of-the art.  We have found that the
contribution from hadronic scattering both to the $\rm R_{AA}$ and to
the elliptic flow of $D$ mesons is appreciable, especially in central
collisions, and produces additional elliptic flow $v_2$ \cite{TSong}.

Since the PHSD results reproduce the experimental data from the STAR
and ALICE Collaborations without radiative energy loss in the $p_T$ range
considered, we conclude that collisional energy loss is dominant at
least up to $p_T=$ 6 (15) GeV/c in relativistic heavy-ion collisions at RHIC (LHC). In our
approach this is essentially due to the infrared enhanced coupling $\alpha_s(T)$ in
the DQPM leading to large scattering cross sections of charm quarks
with partons at temperatures close to $T_c$.

\section*{Acknowledgements}

The authors acknowledge inspiring discussions with J. Aichelin, P. B.
Gossiaux, C. M. Ko, O. Linnyk, R. Marty, V. Ozvenchuk, and R.
Vogt.
This work was supported by DFG under contract BR 4000/3-1, and
by the LOEWE center "HIC for FAIR". JMTR is supported by the Program
TOGETHER from Region Pays de la Loire and the European I3-Hadron
Physics program. LT acknowledges support from the Ramon y Cajal
Research Programme and contracts FPA2010-16963 and FPA2013-43425-P
from Ministerio de Ciencia e Innovaci\'on, as well as from
FP7-PEOPLE-2011-CIG under Contract No. PCIG09-GA-2011-291679. The
computational resources have been provided by the LOEWE-CSC.


\section*{References}
\begin{thebibliography}{9}
\bibitem{lQCD}  Aoki Y {\it et al.} 2006 {\it Phys. Lett.} B {\bf 643} 46
\bibitem{Borsa}  Borsanyi S {\it et al.} 2010
{\it JHEP} {\bf 1009} 073; 2010 {\it JHEP} {\bf 1011} 077; 2012 {\it JHEP}
{\bf 1208} 126; 2014 {\it Phys. Lett.} B {\bf 730} 99; 2015 {\it Phys. Rev.} D {\bf 92} 014505

\bibitem{Peter} Petreczky P [HotQCD Collaboration] 2012
{\it PoS LATTICE}
{ \bf 2012} 069
\bibitem{Ding} Ding H-T Karsch F  Mukherjee S 2015  arXiv:1504.05274


\bibitem{QM2014}
Proceedings of ’Quark Matter-2014’ 2014 {\it Nucl. Phys.} A {\bf 931} 1

\bibitem{Vogt}
 Cacciari M  Nason P  Vogt R 2005 {\it Phys. Rev. Lett.} {\bf 95}  122001

\bibitem{eePHENIX}
 Adare A {\it et al.} (PHENIX Collaboration) 2007 {\it  Phys. Rev. Lett.} {\bf 98}
172301

\bibitem{eeSTAR}
  Abelev B I {\it et al.}  [STAR Collaboration] 2007
{\it  Phys.\ Rev.\ Lett.}  {\bf 98} 192301


\bibitem{Adamczyk:2014uip}
  Adamczyk L {\it et al.}  [STAR Collaboration] 2014
  {\it Phys.\ Rev.\ Lett.}  {\bf 113}  142301

\bibitem{Tlusty:2012ix}
  Tlusty D [STAR Collaboration] 2013
 {\it Nucl.\ Phys.} A {\bf 904-905} 639c

\bibitem{expLHC}   Sakai S {\it et al.} 2013 {\it Nucl. Phys.} A {\bf 904-905} 661c

\bibitem{vanHees:2005wb}
  van Hees H  Greco V and Rapp R
2006 {\it  Phys.\ Rev.} C {\bf 73} 034913

\bibitem{BAMPS}
  Uphoff J {\it et al.} 2013 {\it Nucl. Phys.} A {\bf 910-911} 401;
   {\it ibid} 2014 {\bf 931} 535.

\bibitem{Das:2013kea}
  Das S K Scardina F Plumari S Greco V 2014
 {\it Phys.\ Rev.} C {\bf 90}  044901
\bibitem{Cassing:2008nn}
  Cassing W 2009
 {\it Eur.\ Phys.\ J.\ ST} {\bf 168} 3; 2007 {\it Nucl. Phys.} A {\bf 795} 70

\bibitem{Ozvenchuk:2014rpa}
 Ozvenchuk V {\it et al.} 2014 {\it Phys.\ Rev. } C {\bf 90} 054909

\bibitem{PHSD}
Cassing W Bratkovskaya E L 2009 {\it Nucl. Phys.} A {\bf 831} 215

\bibitem{PHSDrhic}
  Bratkovskaya E L Cassing W Konchakovski V P Linnyk O 2011
 {\it Nucl.\ Phys.} A {\bf 856} 162



\bibitem{Volo}  Konchakovski V P  {\it et al.} 2015 {\it J. Phys.} G {\bf 42} 055106; 2014 {\it J. Phys.} G {\bf 41} 105004; 2014 {\it  Phys. Rev.} C {\bf 90} 014903; 2012
{\it Phys. Rev.} C {\bf 85}  044922; 2012 {\it Phys. Rev.} C {\bf 85} 011902

\bibitem{Linnyk}  Linnyk O {\it et al.} 2014 {\it  Phys. Rev.} C {\bf 89}  034908; 2013 {\it  Phys. Rev.} C {\bf 88} 034904; 2013 {\it  Phys. Rev.} C {\bf 87}   014905; 2012 {\it Phys. Rev.} C {\bf 85} 024910; 2011 {\it Phys. Rev.} C {\bf 84} 054917; 2011 {\it Nucl. Phys.} A {\bf 855} 273
\bibitem{HSD}
      Cassing W Bratkovskaya E L 1999 {\it Phys. Rep.} {\bf 308} 65
\bibitem{HSD2}
  Cassing W Bratkovskaya E L  Juchem S 2000
{\it Nucl. Phys.} A {\bf 674} 249

\bibitem{TSong}  Song T Berrehrah H Cabrera D
Torres-Rincon J M  Tolos L  Cassing W  Bratkovskaya E L 2015
{\it Phys. Rev.} C {\bf 92} 014910

\bibitem{Adamczyk:2012af}
  Adamczyk L {\it et al.}  [STAR Collaboration] 2012
  {\it Phys.\ Rev.} D {\bf 86} 072013

\bibitem{delValle:2011ex}
  Conesa del Valle Z [ALICE Collaboration] 2012
  {\it AIP Conf.\ Proc.}  {\bf 1441} 886
\bibitem{Berrehrah:2013mua}
  Berrehrah H {\it et al.} 2014
 {\it Phys.\ Rev.} C {\bf 89}  054901

\bibitem{Berrehrah:2014kba}
    Berrehrah H {\it et al.} 2014
 {\it Phys.\ Rev.} C {\bf 90}  064906

\bibitem{Younus:2013rja}
  Younus M Coleman-Smith C E Bass S A Srivastava D K 2015
 {\it Phys.\ Rev.} C {\bf 91} 024912

\bibitem{Hamza14}  Berrehrah H {\it et al.} 2014  	
{\it Phys. Rev.} C {\bf 90} 051901


\bibitem{Tolos:2013kva}
  Tolos L Torres-Rincon J M 2013
 {\it Phys.\ Rev.} D {\bf 88} 074019



\bibitem{Shadowing}
 Eskola K J Paukkunen H Salgado C A 2009
{\it JHEP} {\bf 0904} 065


\bibitem{ALICE:2012ab}
  Abelev B {\it et al.} [ALICE Collaboration] 2012
 {\it JHEP} {\bf 1209}  112


\bibitem{ALICE:2013v2}
 Abelev B {\it et al.} [ALICE Collaboration] 2013
{\it Phys. Rev. Lett.} {\bf 111} 102301


\end{thebibliography}
\end{document}